# Towards Representation and Validation of Knowledge in Students' Learning Pathway Using Variability Modeling Technique

Abdelrahman Osman Elfaki, Md Gapar Md Johar, Kevin Loo Teow Aik, Sim Liew Fong, Ruzi Bachok

Faculty of Information Science and Engineering

Management and Science University, Malaysia

abdelrahmanelfaki@gmail.com,{ gapar, Kevin, lfsim & ruzi}@msu.edu.my

*Abstract* - **Nowadays, E-learning system is considered as one of the main pillars in the learning system. Mainly, E-Learning system is designed to serve different types of students. Thus, providing different learning pathways are a must. In this paper, we introduce the variability technique to represent the knowledge in E-learning system. This representation provides different learning pathways which supports the students' diversity. Moreover, we validate the selection of learning pathway by introducing First Order Logic (FOL) rules.**

## I. INTRODUCTION

Learning pathway is defined as the chosen route taken by a learner, which allows him to build knowledge progressively, i.e., in learning pathways, the control of choice moves away from the tutor to the learner [1]. Learning pathway aids learners to access information and courses by which they can construct personalized transitions between the information to be accessed and their own cognitive structures [2]. Students in the learning pathway program will be responsible for their own learning in a self-directed, independent manner, including when, where, what, and how to study. Examinations will be taken when students consider themselves ready [3]. In literature, there are many useful usages of learning pathway, e.g., one of the main importance usages of learning pathways according to Jin [4] is to explore and explain human behaviours during learning processes. More useful usages of learning pathways are discussed in [5].

Nowadays, one of the new major directions in research on web-based educational systems is the notion of adaptability [6]. Adaptability means the system should service different preferences and different students' abilities. Moreover, Carchiolo [7] mentioned the importance of supporting e-learning system by a mechanism that works to provide different pathways for students. According to Bauerova [8] the design of learning pathways requires knowledge management system for representing knowledge to assist students in course selection. Although learning pathway has wide implementations recently, still the standard specification of the learning pathway is missing [9, 10]. As a conclusion, the problem that is discussed here is how to represent learning pathway's knowledge using standard notations in which it supports the diversity.

In this paper, we introduce a variability modeling for representing the learning pathway. Our modeling is based on two layers. The first layer is graphical representation of all courses. The second layer is mathematical representation using First Order Logic (FOL). The second layer is a direct translation of the first layer in which every item in the first layer is represented using FOL predicates in the second layer. Moreover, we introduce FOL rules to validate the course selection process.

## II. BACKGROUND: WHAT IS THE VARIABILITY

Variability is defined as the ability of a system to be modified, updated, or customized to be used in a specific context [11]. Variability identifies the common and variant assets within the specific domain [12]. Pohl et al. [13] suggest that the three following questions be answered in order to define variability: What, Why and How?

What? At this point, the variable item is precisely identified. The variable represents a property of the real world. The definition of the term variability introduces a new term, the 'variability subject'. A variability subject is defined as a variable item of a real world property. The word 'machine' is an example of a real world property. The word 'car' is an example of a variability subject of the real world property 'machine'.

Why? Generally, there could be many reasons for an item or property to vary, for example, diversity of stakeholders' needs, diversity of business and country laws, and technical issues. Furthermore, for related items, the diversity of one item could be the reason for the variation of other related item.

How? To answer this question, the term 'variability object' is introduced. Different shapes can be handled by one variability subject. A variability object is defined as a particular instance of a variability subject. For example, the word 'car' is a variability subject. The words 'Toyota', 'Nissan', and 'BMW' are examples of variability objects of the variability subject 'car'.

## III. REPRESENTING LEARNING PATHWAY'S KNOWLEDGE USING VARIABILITY NOTATIONS

In the learning pathway, different courses can be presented based on the study area (for example, software engineering, or networking). In order to implement the selection of learning pathway, the study area is divided into main points or what we called here as fields. Each field could be divided into options.

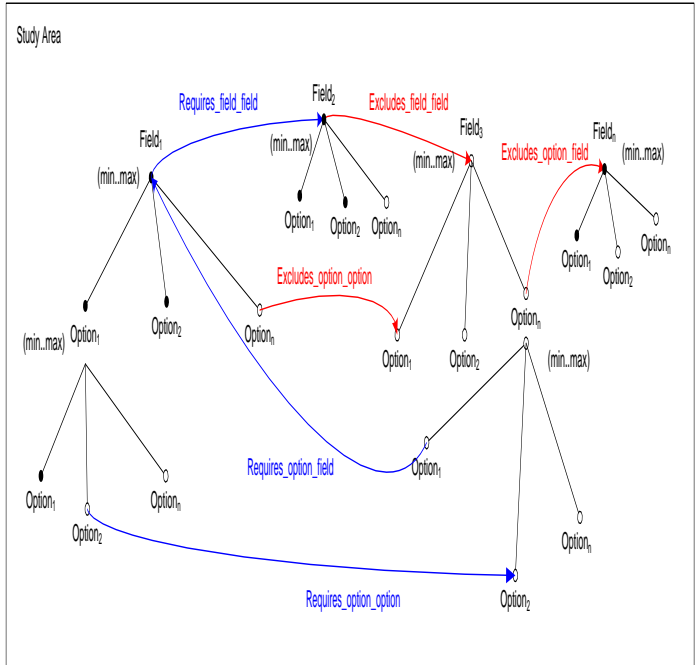

Figure 1: General Structure of our proposal modeling.

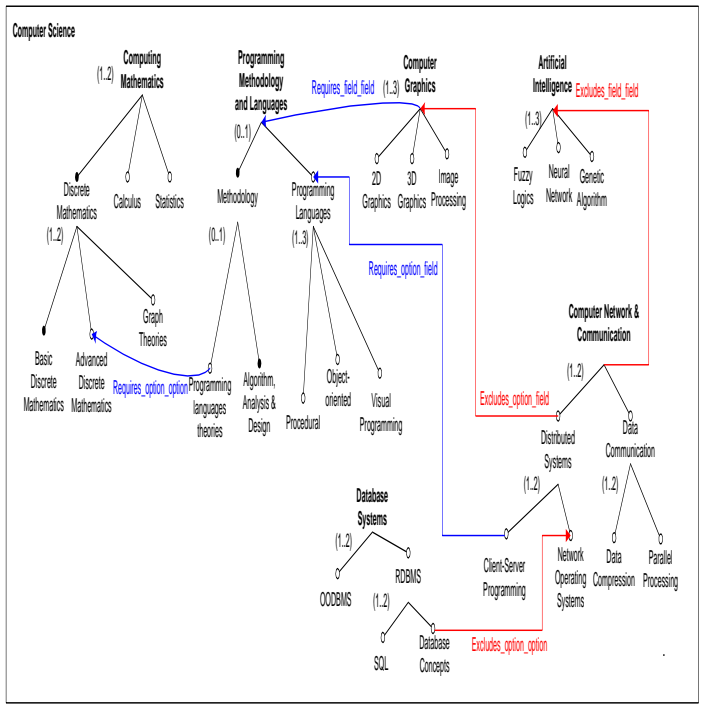

Figure 2: Example of the upper layer of our model

Figure 1, illustrates the general structure of our modeling. Figure 2 shows an example of our modeling. In figure 2, computer science represents the study area; programming, mathematical computing, and artificial intelligence could represent the fields; in programming field: Java, C, and C++ could represent the options. In any field, students must select one option or more. Cardinality describe the minimum and maximum number that must be selected from each field. For instance, in figure 2, the field computing mathematics has the cardinality of 1 to 3. There are two possibilities for the fields and options, either common or not common. Common field means that this field must be chosen to be included in any pathway. In more details, common field means that the student must select one or more from its options. For instance, in figure 2, methodology represents a common field, and programming language theory represents not common option.

According to [13], variability elements are: decision points, choices, and constraint dependency rules between them. Therefore, we can claim that our notations (after we add the constraint dependency rules) are enough to represent the variability.

In the following, the notations of the proposed modeling are explained. Our modeling consists of two layers. The upper layer is graphical notations that work as user interface to guide user among the selection process. The lower layer is a direct translation of the upper layer. First, the upper layer is described and later fields, options, and constraint dependency rules are represented using predicates as a lower layer. The output of the lower layer is a complete modeling of student pathway as a knowledge-base.

### A. The Upper Layer

Nowadays, Feature Model (FM)[14] and Orthogonal Variability Model (OVM) [13] are the well-known techniques to represent variability. In our modeling, we combined FM diagram with OVM notations. The upper layer of our model is a graphical representation to satisfy the visualization condition. Visualization is defined in [15] as a graphical representation in a hierarchical form in order to increase the understandability of the variability.

Our first layer is a merging between FM and OVM notations. Figure 1 represents the example of the upper layer of our modeling. Optional and mandatory constraints are defined in figure 1 by original FM notations [14], and constraint dependency rules are described using OVM notations.

According to Milašinović [16], a graphical representation is a suitable form to illustrate the course structure. Although the graphical representation support readability which enhances the user's interactivity, there is no formal definition for the graphical representations. The lack of formal definition prevents the use of standard software tools to validate the selection process. The lower layer is suggested to provide formal definition for our graphical model, i.e., the upper layer.

*B. The Lower Layer Representation*

The lower layer of the proposed approach is a logical representation of the variability using FOL. Representing variability using FOL provides formalization for the proposed modeling. In the following, field, option, and dependency constraint rules are described using predicates as a lower layer of the proposed modeling (examples are based on figure 2). Terms starting with capital letters represent variables and terms starting with lowercase letters represent constants.

*a) Field*

A field is a point that can select one or more of its options. Five predicates (type, choiceof, max, min, and common) are used to represent each field. In the following, we explain the syntaxes and semantics of these predicates:

**type**:
Syntax: type(Id,field).
Semantic: this predicate is used to define the type of item in the learning path. In the learning path, the item is either a field or an option. The term "Id" identifies the identification of the field and the term "field" is a constant denoting that this item is field. For instance, type( computer graphics, field).

**Choiceof**:
Syntax: choiceof($Id_1$,$Id_2$).
Semantic: identifies the field of a specific option, i.e., assigns each option to its field. $Id_2$ is an option that belongs to the filed $Id_1$. For instance, choiceof(computer graphics,2D garphics). This example shows that "2D graphics" is an option belongs to field "computer graphics".

**max**:
Syntax: max($Id_1$,int)
Semantic: identifies the maximum number allowed to be selected from a specific field. $Id_1$ is identification of a field and *int* is an integer. For instance, max(computer graphics, 3).

**min**:
Syntax: min($Id_1$, int).
Semantic: identifies the minimum number allowed to be selected from a field. $Id_1$ is identification of a field and *int* is an integer. For instance, min(computer graphics, 1).
The common option(s) in a field is/are not included in maximum-minimum numbers of selection.

**common**:
Syntax: common($Id_1$,yes).
Semantic: describes the commonality of the field (whether the field is common or not). $Id_1$ represents identification of the field. The term "yes" is a constant denotes that this field is common. For instance, common(methodology, yes).

The commonality of the item denotes that this item is common for the specific study area and must be included in any student learning pathway.

*b) Option*

Option is a choice that belongs to a specific field. As an example, in figure 2, "2D graphics" is an option belongs to the "computer graphics" field. Two predicates are used to represent each option (type and common). In the following, the syntaxes and semantics of these predicates are illustrated:

type:
Syntax: type(Id1,option).
Semantic: defines the type of feature. Id1 represents the variant name, e.g., type(3D graphics, option).

common:
Syntax: common(Id1,yes).
Semantic: describes the commonality of the option (whether the option is common or not). Id1 represents option's identification, e.g., common(discrete mathematics, yes).

The option could be option and field in the same time, e.g., "discrete mathematics". The option that has some also options is defined as option and field in the same time.

*c) Constraint dependency rules*

The "require" and "exclude" relations are used to describe the dependency constraint [13]. Require means if an item is selected and this item requires another item then the required item must follow the selected item. Exclude means if an item is selected and this item excludes another item then the excluded item must not be selected. Item can be field or option.

Here, we define the six relations to describe the constraint dependency rules discussed in [13], to be implemented in our proposed model. These relations are: option requires another option, option excludes another option, option requires field, option excludes field, field requires another field, and field excludes another field. The relation, field require/exclude another field, is not included in the constraint dependency rules because it could be divided into option require/exclude another option relations.

In our proposed modeling, six predicates are used to describe the constraint dependency rules. In the following, the syntaxes and semantics of these predicates are illustrated:

**requires_option_option**:
Syntax: requires_option_option($Id_1$,$Id_2$).
Semantic: option requires another option. $Id_1$ represents the requiring option and $Id_2$ represents the required option, e.g., requires_option_option(programming language theories, advance discrete mathematics).

**excludes_option_option**:
Syntax: excludes_option_option($Id_1$, $Id_2$).
Semantic: option excludes another option. Id1 represents the exclusion option and $Id_2$ represents the excluded option, e.g., excludes_option_option(database concepts, network operating systems).

**requires_option_field**:
Syntax: requires_option_field($Id_1$,$Id_2$).
Semantic: option requires field. $Id_1$ represents the requiring option and $Id_2$ represents the required field, e.g., requires_option_field(client server programming, programming language).

**excludes_option_field**:
Syntax: excludes_option_field($Id_1$,$Id_2$).

Semantic: option excludes field. $Id_1$ represents the requiring option and $Id_2$ represents the excluded field, e.g., excludes_option_field(distributed systems, computer graphics).

**requires_field_field**:
Syntax: requires_field_field($Id_1$, $Id_2$).
Semantic: field requires another field. $Id_1$ represents the requiring field and $Id_2$ represents the required field, e.g., Requires_field_field(computer graphics, programming methodology and language)

**excludes_field_field**:
Syntax: excludes_field_field($Id_1$, $Id_2$).
Semantic: field excludes another field. $Id_1$ represents the exclusion field and $Id_2$ represents the excluded field, e.g., excludes_field_field (computer network and communication, Artificial Intelligence).

*d) Additional Predicates*

In our proposed modeling, we use additional predicates to complete the validation operations. In the following, each additional predicate is defined:

**select:**
Syntax: select($Id_i$).
Semantic: the predicates select($Id_i$) mean that the item $Id_i$ is selected by the user. Item could be either field or option.

**notselect:**
Syntax: notselect($Id_i$).
Semantic: the predicates notselect($Id_i$) mean that the item $Id_i$ is excluded from being part of the student learning pathway.

**no-selected:**
Syntax: no-selected($Id_i$,n).
Semantic: the predicate no-selected($Id_i$,n) counts the number of selected items ($Id_i$) in the student learning pathway. The letter n is an integer denoting the number of selected $Id_i$.

Table1 and Table 2 illustrate lower layer representation. Table 1 shows the representation of the filed "computer graphics". Table 2 illustrates representation of the option "distributed systems".

TABLE 1
REPRESENTATION OF THE FIELD COMPUTER GRAPHICS IN THE LOWER LAYER

| |
|---|
| type(computer graphics, field). |
| choiceof(computer graphics,2D graphics). |
| choiceof(computer graphics,3D graphics). |
| choiceof(computer graphics, image processing). |
| requires_field_field(computer graphics, programming methodology and languages). |
| common(computer graphics, no). |
| min(computer graphics,1). |
| max(computer graphics,3). |

TABLE 2
REPRESENTATION OF THE OPTION DISTRIBUTED SYSTEM IN THE LOWER LAYER

| |
|---|
| type(distributed systems, option). |
| common(distributed systems, no). |
| excludes_option_field(distributed systems, computer graphics). |

## IV. VALIDATING THE SELECTION OF STUDENT LEARNING PATHWAY

By validating the selections of students' learning pathway, we mean the checking of the satisfaction of four constraints: constraint dependency rules (require and exclude), the relation between the field and its option, commonality (whether the item is common or not), and cardinality of the field. The proposed modeling triggers rules for validating the following constraints:

1) Constraint dependency rules. According to Pohl et al [13], there are six constraint dependency rules;
2) The relation between the field and its options and vice versa. As an example, if a field is selected, this means that its common options are also selected;
3) Commonality. If the field is common, this means that it must be included in any student pathway; and
4) Cardinality: Cardinality defines the maximum and minimum numbers allowed to be selected from the any field.

TABLE 3
THE GENERAL VALIDATION RULES IN THE USER'S SELECTIONS

| |
|---|
| **Definitions:** type($Id_1$,field),type($Id_2$,field),type($Id_3$,option),type($Id_4$,option), choiceof($Id_1$,$Id_4$), and n is an integer. |
| $\forall Id_3, Id_4$: require_option_option($Id_3$, $Id_4$) $\wedge$ select($Id_3$) $\Rightarrow$ select($Id_4$)   (1) |
| $\forall Id_3, Id_4$: exclude_option_option($Id_3$, $Id_4$) $\wedge$ select($Id_3$) $\Rightarrow$ notselect($Id_4$)   (2) |
| $\forall Id_3, Id_1$: require_option_field($Id_3$, $Id_1$) $\wedge$ select($Id_3$) $\Rightarrow$ select($Id_1$)   (3) |
| $\forall Id_3, Id_1$: exclude_option_field($Id_3$, $Id_1$) $\wedge$ select($Id_3$) $\Rightarrow$ notselect($Id_1$)   (4) |
| $\forall Id_1, Id_2$: require_field_field($Id_1$, $Id_2$) $\wedge$ select($Id_1$) $\Rightarrow$ select($Id_2$)   (5) |
| $\forall Id_1, Id_2$: exclude_field_field($Id_1$, $Id_2$) $\wedge$ select($Id_1$) $\Rightarrow$ notselect($Id_2$)   (6) |
| $\forall Id_4, Id_1$: select($Id_4$) $\Rightarrow$ select($Id_1$)   (7) |
| $\exists Id_4, \forall Id_1$: select($Id_1$) $\Rightarrow$ select($Id_4$)   (8) |
| $\forall Id_4, Id_1$: notselect($Id_1$) $\Rightarrow$ notselect($Id_4$)   (9) |
| $\forall Id_4, Id_1$: common($Id_4$,yes) $\wedge$ select($Id_1$) $\Rightarrow$ select($Id_4$)   (10) |
| $\forall Id_1$: common($Id_1$,yes) $\Rightarrow$ select($Id_1$)   (11) |
| $\forall Id_4, Id_1$: select($Id_4$) $\wedge$ (no_selected($Id_4$,n) $\geq$ max($Id_1$,n)) $\Rightarrow$ notselect($Id_4$) (12) |
| $\forall Id_4, Id_1$: select($Id_4$) $\wedge$ (no_selected($Id_4$,n) $\leq$ min($Id_1$,n)) $\Rightarrow$ notselect($Id_4$) (13) |

Each of the rules in Table 3 is described in the following:
**Rule 1**:
For all options $Id_3$ and $Id_4$; if $Id_3$ requires $Id_4$ and $Id_3$ is selected, then $Id_4$ is selected.
**Rule 2:**
For all options $Id_3$ and $Id_4$; if $Id_3$ excludes $Id_4$ and $Id_3$ is selected, then the *notselect* predicate is assigned to $Id_4$.
**Rule 3**:
For all options $Id_3$ and field $Id_1$; if $Id_3$ requires $Id_1$ and $Id_3$ is selected, then $Id_1$ is selected.
**Rule 4**:
For all options $id_3$ and field $Id_1$; if $Id_3$ excludes $Id_1$ and $Id_3$ is selected, then the *notselect* predicate is assigned to $Id_1$.
This rule is also applicable if the field is selected first:

∀ $Id_3$, $Id_1$: $type(Id_3, option) \land type(Id_1, field) \land exclude\_option\_field(Id_3, Id_1) \land select(Id_1) \Rightarrow notselect(Id_3)$, i.e., for all option $Id_3$ and filed $Id_1$; if $Id_3$ excludes $Id_1$ and $Id_1$ is selected, then the *notselect* predicate is assigned to $Id_3$.

**Rule 5**:
For all field $Id_1$ and field $Id_2$; if $Id_1$ requires $Id_2$ and $Id_1$ is selected, then $Id_2$ is selected.

**Rule 6:**
For all field Id1 and Id2; if Id1 excludes Id2 and Id1 is selected, then the notselect predicate is assigned to Id2.

**Rule 7:**
For all option Id4 and field Id1; where Id4 belongs to Id1 and Id4 is selected, this means that Id1 is selected. This rule determines the selection of the field if one of its options has been selected.

**Rule 8**:
For all fields Id1 there exists an option Id4; if Id1 is selected and Id4 belongs to Id1, then Id4 is selected. This rule states that if a field has been selected, then the option(s) belonging to this field must be selected.

**Rule 9**:
For all option Id4 and field Id1; where Id4 belongs to Id1 and the predicate notselect is assigned to Id1, then the *notselect* predicate is assigned to Id4 also. This rule states that if a field has been excluded, then none of its options must be selected.

**Rule 10**:
For all option Id4 and field Id1; where Id4 is a common option and belongs to Id1 and Id1 is selected, then Id4 is selected. This rule states that if an option is common and its field is selected, then this option must be selected in all student learning pathway.

**Rule 11**:
For all field Id1; if Id1 is common, then Id1 is selected. This rule states that if a field is common then it must be selected in the learning pathway.

**Rule 12**:
For all option Id4 and field Id1; where Id4 belongs to Id1 and Id4 is selected, then the summation of Id4 must be less than the maximum number that is allowed to be selected from Id1.

**Rule 13**:
For all option Id4 and field Id1; where Id4 belongs to Id1 and Id4 is selected, then the summation of Id4 must be greater than the minimum number allowed to be selected from Id1. The *notselect* predicate prevents any option or any field from being selected.

## V. RELATED WORKS

In the literature, many researchers focused on cognitive process when they study the students' learning pathways. In this paper, we focus in the real and physical selection of the course materials. Mainly, this is a hot issue in E-learning systems. Semet et al.[17] optimize the searching in students' pathway using Ant Colony Optimization (ACO). In Semet et al.[17] model, the underlying structure of the E-learning material is represented by a graph with valued arcs whose weights are optimized by virtual ants that release virtual pheromones along their paths. This gradual modification of the graph's structure improves learning pathways by matching similar topics. Carchiolo et al. [18] provide adaptive environment for E-learning system by defining different learning pathways and match between students' capabilities and suitable pathway. Chen et al.[19] introduce association rules to tune the learning pathway. The model in [18] mined the learners' profile to discover the common misconception and after that tune courseware structure through modifying the difficulty parameters of courseware in the courseware database. Wong and Looi [20] suggest ACO as a tool to adapt the learning pathway. Our modeling can provide all the functions that are discussed in [17-20]. To the best of our knowledge, our work is the first work that introduces the variability as a modeling technique in e-learning systems.

## VI. CONCLUSION

In reality, different types of students with different backgrounds are using the same E-learning system. Therefore, modeling E-learning system to allow different students' pathways is a must. In this paper, we introduce a new model for modeling students' learning pathway. In this proposed modeling, variability is explicitly represented which satisfy the diversity.

In our model, the E-learning system is divided into group of study areas. Each study area is also divided into groups of fields, and each field is divided into groups of options. Our model consists of two layers. The first layer is graphical interface, and the second layer is mathematical representation of the first one. In the second layer, fields and options are represented using FOL predicates. Moreover, we introduce FOL rules to validate the students' learning pathway selection.

In our future work, we plan to implement this model in our E-learning system of http://klas.msu.edu.my/login.php.